\documentclass[11pt]{article} 
\usepackage[tbtags]{amsmath}   
\usepackage{multicol}
\usepackage{amssymb}     
\usepackage{relsize}
\usepackage{array}
\usepackage[margin=0.75in]{geometry}    
\usepackage{graphicx}   
\usepackage{tabularx}  
\usepackage[square,     
            numbers,      
            sort&compress 
            ]{natbib}
\usepackage{setspace}
\usepackage{xcolor}
\usepackage{caption}
\usepackage{placeins}
\usepackage{afterpage}
\newcolumntype{P}[1]{>{\centering\arraybackslash}p{#1}} 

\begin{document}

\centerline{\LARGE\bf{Flame Instability and Transition to Detonation}}
\centerline{\LARGE\bf{in Supersonic Reactive Flows}}
\bigskip
\centerline{\large\it{Gabriel B. Goodwin}}
\medskip
\centerline{Naval Center for Space Technology}
\centerline{Naval Research Laboratory}
\centerline{4555 Overlook Ave. SW, Washington, DC, USA 20375} 
\centerline{Corresponding author: gabe.goodwin@nrl.navy.mil}
\bigskip
\centerline{\large\it{Elaine S. Oran}}
\medskip
\centerline{Department of Aerospace Engineering}
\centerline{University of Maryland, College Park}

\doublespacing
\section*{Abstract}

Multidimensional numerical simulations of a homogeneous, chemically reactive gas were used to study ignition, flame stability, and deflagration-to-detonation transition (DDT) in a supersonic combustor. The configuration studied was a rectangular channel with a supersonic inflow of stoichiometric ethylene-oxygen and a transimissive outflow boundary. The calculation is initialized with a velocity in the computational domain equal to that of the inflow, which is held constant for the duration of the calculation. The compressible reactive Navier-Stokes equations were solved by a high-order numerical algorithm on an adapting mesh. This paper describes two calculations, one with a Mach 3 inflow and one with Mach 5.25. In the Mach 3 case, the fuel-oxidizer mixture does not ignite and the flow reaches a steady-state oblique shock train structure. In the Mach 5.25 case, ignition occurs in the boundary layers and the flame front becomes unstable due to a Rayleigh-Taylor instability at the interface between the burned and unburned gas. Growth of the reaction front and expansion of the burned gas compress and preheat the unburned gas. DDT occurs in several locations, initiating both at the flame front and in the unburned gas, due to an energy-focusing mechanism. The growth of the flame instability that leads to DDT is analyzed using the Atwood number parameter. 

\noindent
{\sl Keywords:} Turbulent flame; Hypersonics; DDT; Numerical simulations

\section{Introduction}

A comprehensive understanding of combustion in high-speed flows is necessary for the development of reliable hypersonic vehicles. Airbreathing engines that operate at high supersonic and hypersonic flight speeds will enable the next generation of extended-range, rapid response missile systems and low-cost space access due to significant gains in efficiency over traditional rocket engines \cite{waltrup2002history,daines1998combined}. Maintaining flame stability across a wide range of inflow conditions in the combustor of an airbreathing hypersonic vehicle is a challenge \cite{rasmussen2005stability,ben2001cavity} due to the small timescales for mixing, ignition, and complete combustion. The fundamental physics underlying ignition and combustion in high-speed flows is complex due to the shocks, boundary layers, turbulence, and chemical reactions present in these systems. Experimental and computational studies exploring the interaction of these phenomena will enable the design of higher-performance hypersonic engines.

Combustion of premixed fuels and the influence of high-speed turbulence on flame development has been studied computationally \cite{poludnenko2010interaction,hamlington2011interactions,poludnenko2011interaction}, finding that under certain regimes, premixed turbulent flames are inherently unstable \cite{poludnenko2015pulsating}. Detonation waves may be used instead of a flame or deflagration as a propulsion mechanism in a hypersonic engine \cite{kailasanath2000review,kailasanath2000performance}. Detonation engines have been studied computationally \cite{schwer2011numerical,schwer2013fluid,tsuboi2015numerical} and experimentally \cite{allgood2006performance,dyer2012parametric,lefkowitz2015schlieren} to characterize performance and detonation wave stability under a range of conditions. Prior work has examined the mechanisms for the deflagration-to-detonation transition (DDT) \cite{OranCNFrev2007,goodwin2017shock,gamezoPCI2007}, which is required for ignition of a detonation engine. Hypersonic combustor test facilities are used to study the stabilization of detonation waves in premixed flows \cite{sosa2017advances}.

The purpose of this work is to characterize the effect of high supersonic and hypersonic flow speeds on premixed flames and identify the conditions that lead to flame instability and eventual DDT. As a background to this paper, a series of simulations was performed to investigate the effect of varying inflow Mach number, $M_\infty$, of premixed fuel-oxidizer into a constant-area combustor on ignition, flame growth, formation of fluid instabilities, and DDT. The detailed results of these simulations, for $M_\infty$ = 3 to 10, will be presented in a subsequent paper. The focus of this paper is on two of these cases, $M_\infty = 3$ and 5.25. In the $M_\infty$ = 3 case, the temperature of the fuel-oxidizer mixture does not reach the threshold for autoignition. No combustion occurs and the flow reaches a steady-state repeating oblique shock structure. In the $M_\infty$ = 5.25 case, autoignition occurs in the boundary layer. The flame front becomes turbulent and unstable through shock interactions and growth of fluid instabilities. Eventually, DDT occurs as shocks focus energy at the flame front resulting in the transition to detonation. 

\section{Numerical and Physical Model}

The numerical model solves the full set of Navier-Stokes equations for an unsteady, fully compressible, chemically reacting gas, as described in \cite{OranCNFrev2007}. The reaction of a stoichiometric mixture of ethylene and oxygen is modeled using a simplified, calibrated chemical-diffusive model of the form $dY/dt \equiv \dot{w} = -A\rho Y\exp(-E_\mathrm{a}/RT)$ where $\rho$, $T$, $Y$, and $\dot{w}$ are the mass density, temperature, mass fraction of reactant, and reaction rate, respectively. $A$ is the pre-exponential factor and $E_\mathrm{a}$ is the activation energy. Input parameters for a stoichiometric ethylene-oxygen mixture initially at 298 K and 1 atm, as detailed in \cite{GoodwinCNF2016}, were used. A genetic algorithm optimization procedure \cite{kaplan2017chemical} was used to identify the input parameters that most accurately reproduce the flame and detonation properties for the fuel-oxidizer mixture. This reaction model quantitatively reproduces flame acceleration, onset of turbulence, and DDT mechanisms seen in experiments \cite{xiaoflame,xiao2015formation} and has recently been used to study DDT \cite{GoodwinCNF2016,goodwin2017shock,houim2016role} and layered detonations \cite{houim2017influence}. A Godunov algorithm, fifth-order accurate in space and third-order accurate in time \cite{houimJCP2011}, is used to solve the equations on a dynamically adapting grid \cite{BoxLib}. The simulations described below used grids with minimum size $dx_\mathrm{min}$ = 3.3 $\mu$m and coarsest size $dx_\mathrm{max} = 53.3$ $\mu$m. This choice of grids was tested and shown to resolve the shocks, boundary layers, flames, and other important flow and chemical structures.

The geometrical setup is shown in Fig.~\ref{fig:M3}. A supersonic inflow condition is applied at the left boundary. For both $M_\infty$ = 3 and 5.25, inflow temperature, pressure, and mass fraction of reactant are the same (298 K, 1 atm, 1, respectively) and are constant for the duration of the calculation. The calculation is initiated with velocity everywhere in the domain equal to the velocity at the inflow. The right boundary of the domain is a transmissive, non-reflecting outflow. The top and bottom walls are no-slip adiabatic boundaries. The channel height is 0.32 cm and the length is 3.2 cm. A height of 0.32 cm was determined to be sufficient for capturing the flame and detonation physics of interest. The small scale flame and detonation properties of this fuel-oxidizer mixture, such as fast reactions and energy release, make ethylene-oxygen an ideal fuel-oxidizer mixture for studying ignition and flame development in high-speed premixed reactive flows.




\section{Flame Growth and Transition to Detonation} \label{Main}

\begin{figure}[h!]
	\centering
	\includegraphics[width=144mm]{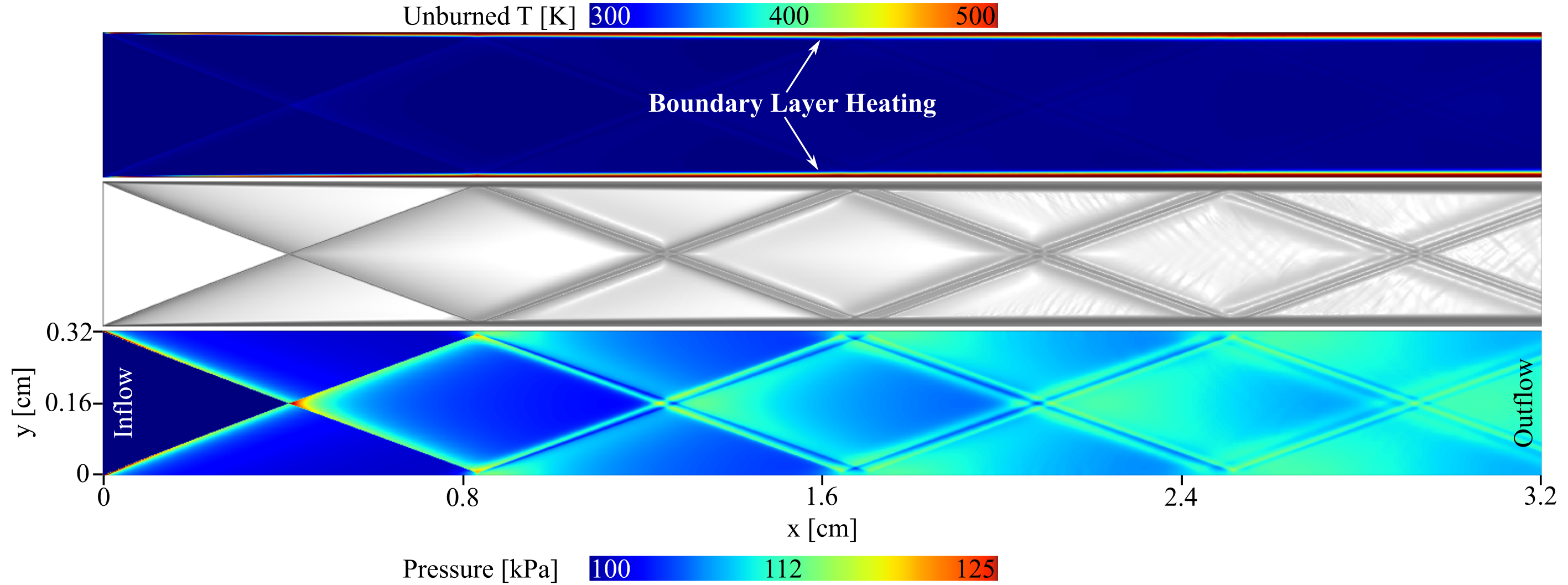}
	\caption{Steady-state solution for Mach 3 inflow. Contours of (top) temperature (middle) numerical schlieren and (bottom) pressure.}
	\label{fig:M3}
\end{figure}

The steady-state flowfield for the $M_\infty$ = 3 case is shown in Fig.~\ref{fig:M3}. The time required to reach steady-state is 50 $\mathrm{\mu s}$. Boundary layers form quickly at the top and bottom walls. The inflow is deflected by the boundary layers, forming oblique shocks that reflect from the wall at a constant angle, a function of $M_\infty$, repeating to the end of the domain. Following the first reflection, the shocks are trailed by expansion waves. Pressure increases as the flow passes through the oblique shock train, resulting in a 10\% increase in average pressure from the inflow to outflow boundaries. Boundary layer temperature reaches a maximum of $\sim$500 K, below the threshold for ignition of the fuel-oxidizer mixture. No combustion occurs and the steady-state oblique shock train remains stationary in the lab reference frame.

\begin{figure}[h!]
	\centering
	\includegraphics[width=144mm]{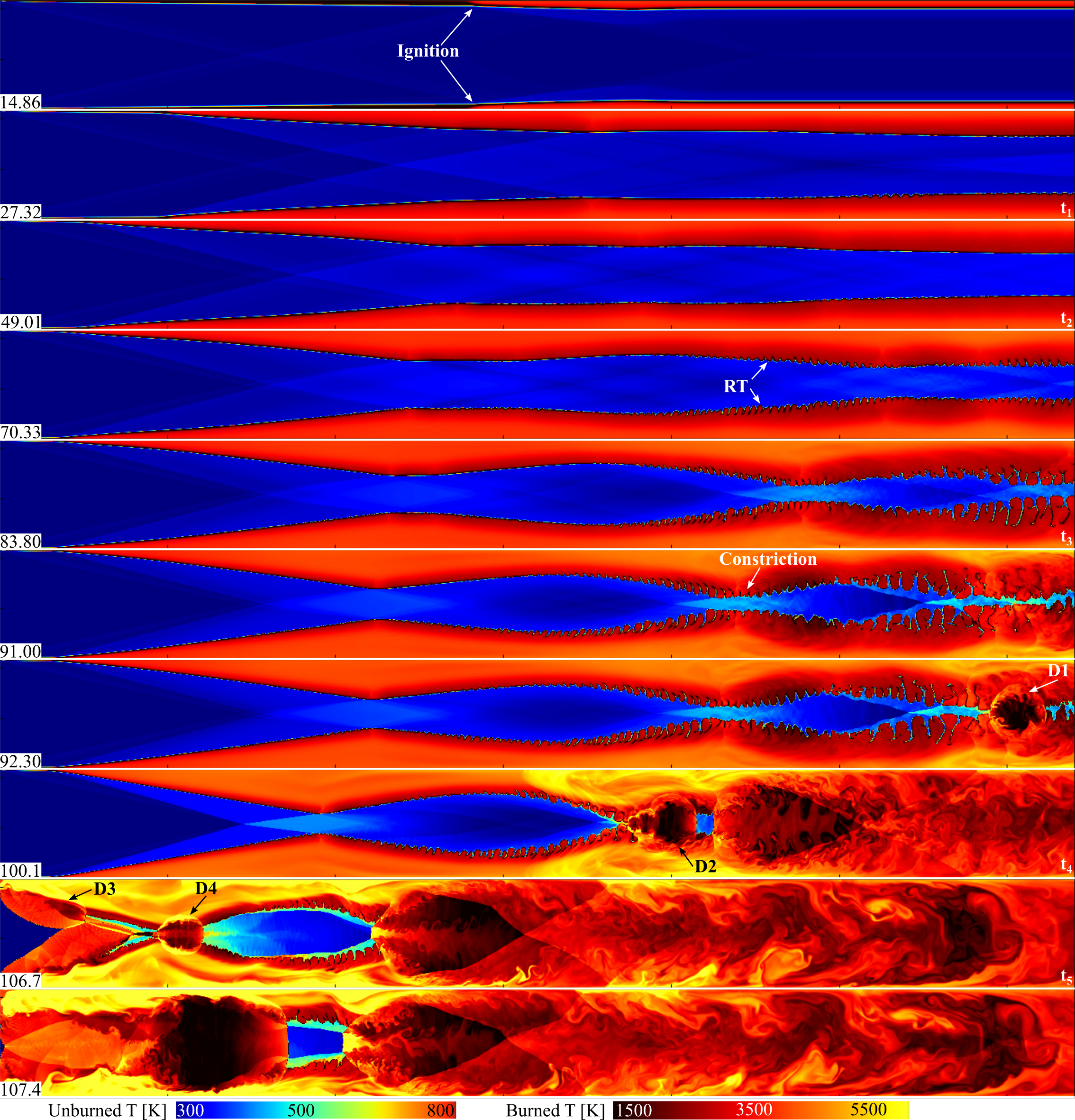}
	\caption{Boundary layer ignition and expansion, growth of Rayleigh-Taylor instabilities (labeled $RT$), and detonation for Mach 5.25 inflow. Time in microseconds shown in frame lower left corners. Entire domain shown.}
	\label{fig:M5+}
\end{figure}

Figure~\ref{fig:M5+} shows temperature contours for the $M_\infty$ = 5.25 case. At 14 $\mathrm{\mu s}$ the fuel-oxidizer mixture ignites in the boundary layer. The initial ignition location is marked in Fig.~\ref{fig:M5+} at 14.86 $\mathrm{\mu s}$. The mixture ignites where the first oblique shock reflects from the channel wall. After ignition, the boundary layers are comprised of burned product that expands into the domain and upstream toward the inflow boundary. An oblique shock train, similar to what is observed for $M_\infty = 3$, forms in the channel of unburned gas between the reaction fronts. The interaction of the shocks with the reaction front results in perturbations to the flame that grow in time due to a Rayleigh-Taylor (RT) fluid instability (detailed in Sec.~\ref{instability}). Bubble and spike structures \cite{Sharp1984} form on the flame surface (labeled RT at 70.33 $\mathrm{\mu s}$ in Fig.~\ref{fig:M5+}). As time progresses, the burned gas expands and the reaction fronts extend further into the channel and closer to the inflow boundary. Intersection of oblique shocks with the flame front creates regions of high temperature and pressure that promote rapid flame expansion at several constriction points (labeled at 91 $\mathrm{\mu s}$).

\begin{figure}[h!]
	\centering
	\includegraphics[width=144mm]{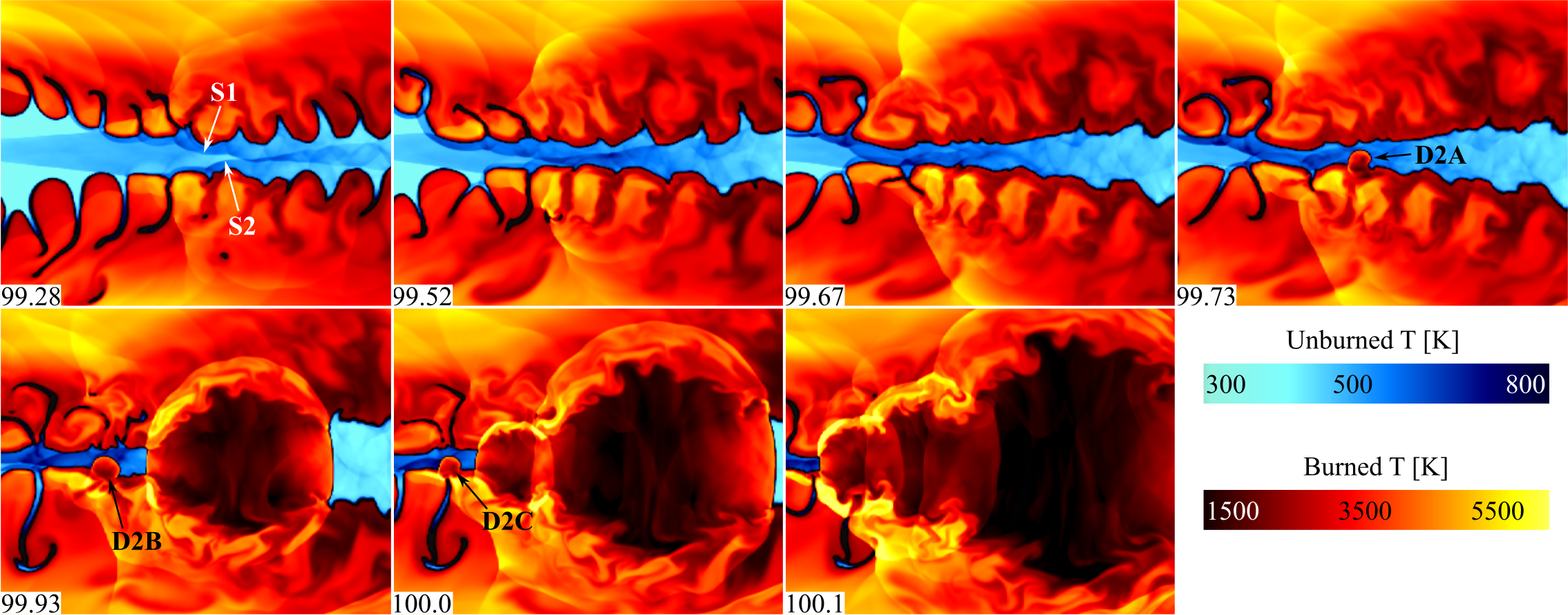}
	\caption{Unstable flame front, shock collision, and detonation. Time in microseconds shown in frame corners. Domain is x = [1.85,2.10] cm and y = [0.075,0.225] cm.}
	\label{fig:M5+_det}
\end{figure}

Temperature and pressure of the unburned gas increase dramatically as it flows through the constricted regions between reaction fronts. Eventually, a detonation is initiated (labeled D1 in Fig.~\ref{fig:M5+}) as a shock passes through the flame front and causes it to transition to a detonation front. The detonation front propagates upstream, consuming the unburned gas flowing between the flame fronts. A subsequent series of detonations is initiated (labeled D2 in Fig.~\ref{fig:M5+}) through the same mechanism as D1. The sequence in Fig.~\ref{fig:M5+_det} is an enlargement of the events leading to D2. Shocks generated by D1 (labeled S1 and S2 in Fig.~\ref{fig:M5+_det}) travel upstream (toward the inflow boundary) ahead of D1 and pass through the flame fronts at D2. The shock amplifies energy release at the flame front, causing it to transition to detonation. The flame transitions to detonation through this mechanism in three locations (labeled D2A, D2B, and D2C in Fig.~\ref{fig:M5+_det}). Transition of a premixed flame to detonation as a result of energy addition of a passing shock has been observed in previous work \cite{GoodwinCNF2016}.

Following the series of detonations at D2, the detonation front formed by D2C travels upstream toward the inflow boundary. Another series of detonations (labeled D3 in Fig.~\ref{fig:M5+}) occurs 5 $\mathrm{\mu s}$ following D2. The mechanism for D3 is the same as that of D2. Shocks generated by D2 pass through the reaction fronts near the inflow boundary and cause the flame to transition to detonation. Immediately following initiation of D3, a final series of detonations (labeled D4 in Fig.~\ref{fig:M5+}) is initiated in the unburned gas between the flame fronts. In D4, the unburned gas is compressed and heated by shocks generated by D2. As a result, the unburned gas detonates through an energy focusing mechanism \cite{goodwin2017shock,GoodwinCNF2016} similar to the mechanism for D1-D3, but in D4 the detonation initiates in unburned gas rather than at a flame front.    

\begin{figure}[h!]
	\centering
	\includegraphics[width=67mm]{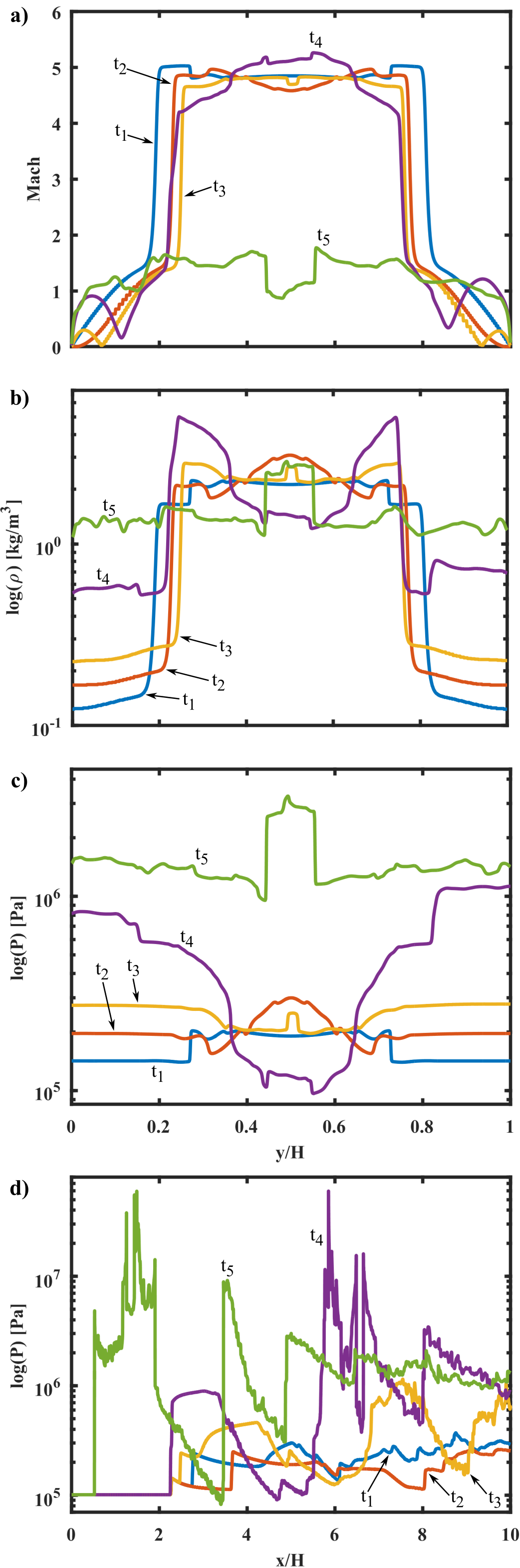}
	\caption{(a) Mach number (b) density (c) pressure across vertical slice through domain at x = 1.6 cm. (d) Pressure across horizontal slice at y = 0.16 cm. Plots (a-c) share common x-axis.}
	\label{fig:line_plots}
\end{figure}

Figure~\ref{fig:line_plots}(a-c) shows profiles of Mach number, density, and pressure at a vertical slice across the channel height at x = 1.6 cm for five timesteps labeled $t_1 - t_5$ correlating to 27.32, 49.01, 83.8, 100.1, and 106.7 $\mathrm{\mu s}$, respectively. These timesteps are shown and labeled in Fig.~\ref{fig:M5+}. The slice location at x = 1.6 cm is labeled $x_2$ in Fig.~\ref{fig:M5+_sch}. Figure~\ref{fig:line_plots}d shows pressure at a horizontal slice across the channel length at y = 0.16 cm for the same timesteps. 

Mach number is greater in the unburned gas than in the burned gas due to the significantly lower temperature and speed of sound. The flow in the burned gas is predominantly subsonic due to the higher speed of sound in this region. Density follows a similar trend. There is a sharp gradient across the flame; density in the burned gas is nearly an order of magnitude lower than density in the unburned gas. Pressure fluctuates in the burned and unburned gas as oblique shocks and expansion waves pass through slice $x_2$. Pressure increases to $>$ 1 MPa at $t_5$ after detonation D2 passes through $x_2$. Figure~\ref{fig:line_plots}(d) shows spikes in pressure through the oblique shock train followed by the pressure drops through the trailing expansion waves. The pressure spikes move closer to the inflow boundary at $x/H$ = 0 as time progresses. 


\section{Stability Analysis} \label{instability}

\begin{figure}[h!]
	\centering
	\includegraphics[width=144mm]{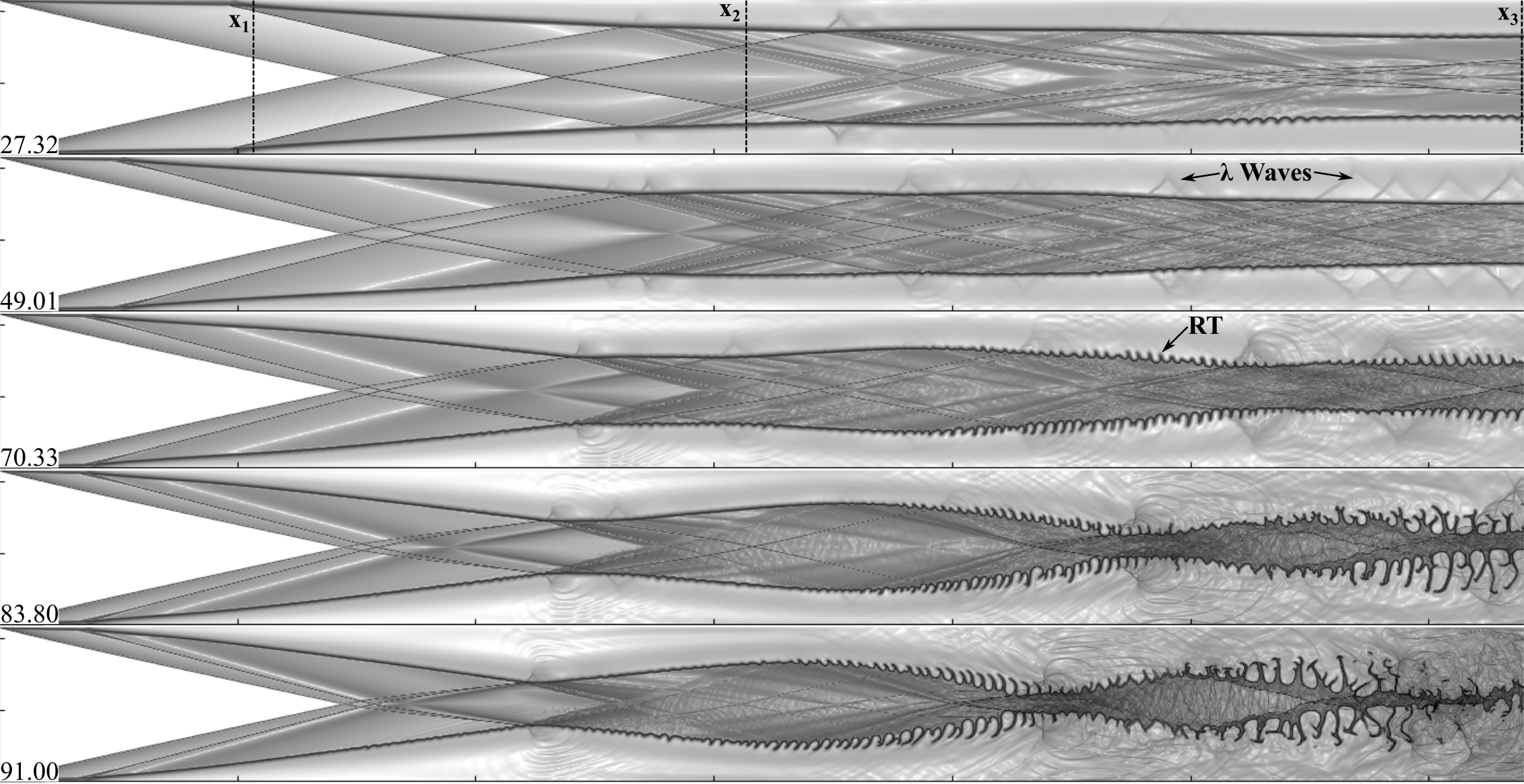}
	\caption{Numerical schlieren for Mach 5.25 inflow. Time in microseconds shown in frame corners. Entire domain shown.}
	\label{fig:M5+_sch}
\end{figure}

For the $M_\infty$ = 5.25 case, the flame front becomes increasingly unstable and turbulent with time. Figure~\ref{fig:M5+_sch} shows a numerical schlieren for the $M_\infty$ = 5.25 calculation. At 27.32 $\mathrm{\mu s}$, the fuel-oxidizer mixture has already ignited in the boundary layers and the flame fronts begin to expand into the center of the channel. An oblique shock train forms in the unburned gas. The oblique shocks pass through the burned gas and reflect from the channel walls generating bifurcated waves, or $\lambda$-shocks (labeled at 49.01 $\mathrm{\mu s}$). These bifurcated waves remain anchored to the points at which the oblique shocks and flame front intersect. Ripples form on the flame surface at these locations. These ripples grow time due to a RT instability at the flame front. The RT instability forms at an interface between two fluids of different densities when the light fluid pushes the heavy fluid \cite{taylor1950instability}. In this case, the low-density burned gas compresses the high-density unburned gas as the reaction front expands into the channel. The height of the ripples increases with time, forming bubble and spike structures (labeled RT in Figs.~\ref{fig:M5+} and \ref{fig:M5+_sch}) typical to the RT instability \cite{Sharp1984}. Bubbles of low-density burned gas push into the high-density unburned gas and spikes of unburned gas form between the bubbles. At 83.8 $\mathrm{\mu s}$, the bubbles have extended far into the center of the channel and nearly the entire channel cross section consists of burned gas. At 91 $\mathrm{\mu s}$, compression of the unburned gas in the center of the channel increases its temperature to $\sim$750 K. Transition to detonation occurs shortly thereafter at D1 as described in Sec.~\ref{Main}.

\begin{figure}[h!]
	\centering
	\includegraphics[width=67mm]{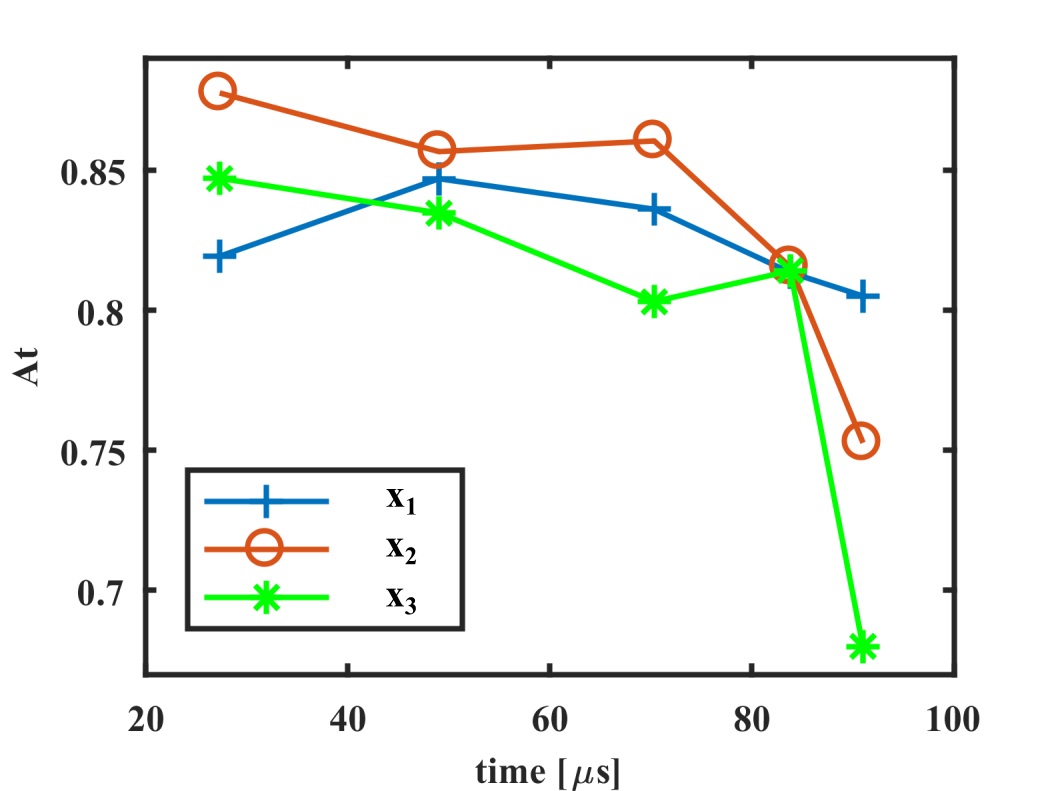}
	\caption{Atwood number as a function of time at three vertical slices through domain.}
	\label{fig:Atwood}
\end{figure}

One of the key parameters used to characterize the growth rate of a RT instability is the Atwood number, $At \equiv (\rho_H - \rho_L)(\rho_H + \rho_L)$, where $\rho_H$ is the density of the heavy fluid and $\rho_L$ is the density of the light fluid. A RT instability is likely to develop for flows with $At < 1$ when the fluids are accelerated in a direction transverse to the interface between them \cite{Sharp1984}. Figure~\ref{fig:Atwood} shows $At$ at three vertical slices through the domain (labeled $x_1$, $x_2$, and $x_3$ in Fig.~\ref{fig:M5+_sch}) for the five timesteps shown in Fig.~\ref{fig:M5+_sch}. The Atwood number was calculated by taking the average unburned gas density across the slice as $\rho_H$ and the average burned gas density as $\rho_L$. The bubbles reach a maximum amplitude at $x_3$, where the lowest $At$ is observed. Amplitude of the bubbles increases drastically from 27.32 $\mathrm{\mu s}$ to 91 $\mathrm{\mu s}$, as shown in Fig.~\ref{fig:M5+_sch}. Additionally, amplitude of the bubbles increases from the inflow to outflow indicating that the flame front becomes more unstable as distance downstream from the point of ignition increases. Structure develops on the spikes due to a Kelvin-Helmholtz instability, or gradient in velocity across the fluid interface, causing the spikes to deform and mushroom \cite{Sharp1984,banerjee2010detailed}. This spike distortion effect is more pronounced at low $At$ \cite{Sharp1984}, particularly for compressible flows \cite{glimm1990numerical}, and is most noticeable near $x_3$ at later timesteps where the lowest $At$ is measured in this study. As seen in the evolution of the flame surface, the RT instability dominates the development of the reaction fronts and causes the formation of bubble-spike structure that compresses and preheats the unburned gas creating ideal conditions for a detonation to occur. 

\section{Solution Sensitivity and Dimensionality}

An additional simulation was performed for the $M_\infty = 5.25$ case with a coarser mesh resolution ($dx_\mathrm{min} = 13.3$ $\mu m$) to investigate the effect of initial conditions on time and location of detonation, flame growth, fluid instabilities, and detonation mechanism. Time and location of the first detonation changed by 10 $\mathrm{\mu s}$ and 0.1 cm, respectively, when compared to the original simulation. Another simulation was performed using a shorter domain in the x direction (2.56 cm) with the same mesh resolution as in the original $M_\infty$ = 5.25 simulation. The first detonation occurs in the same location as detonation D2 in the original simulation with a difference in time to detonation of 9 $\mathrm{\mu s}$. For the shorter domain case, the D1 location in the original simulation would have been outside of the computational domain. In all cases, the same macroscopic flow features are observed: ignition of the fuel-oxidizer mixture in the boundary layer, growth of a RT instability at the flame front, and detonation of the unburned gas due to an energy-focusing mechanism after significant compression and preheating. Other than small deviations in time and location of detonation, the solution is insensitive to domain size or grid resolution. A three-dimensional (3D) simulation was performed to examine the effect of dimensionality on the growth of the flame instability and detonation. The results match what is seen in the 2D $M_\infty$ = 5.25 case with respect to flame evolution and detonation mechanism. A detailed analysis of the 3D simulation will be provided in a subsequent paper.   

\section{Conclusions}

Multidimensional, unsteady numerical simulations of premixed ethylene and oxygen in a supersonic combustor were performed to characterize the effect of flow velocity on ignition, flame expansion and instability, and transition to detonation. This paper examined two cases, $M_\infty$ = 3 and 5.25.

For $M_\infty$ = 3, the fuel-oxidizer mixture does not reach sufficient temperature for ignition and no combustion occurs. In the case with $M_\infty$ = 5.25, the temperature in the boundary layers at the channel walls is sufficient for ignition. The flames in the boundary layers expand and an oblique shock train forms between the reaction fronts. Interaction of the shock and the reaction front results in perturbations that are initially small ripples in the flame. These ripples grow in time due to a Rayleigh-Taylor fluid instability to form bubble and spike structures along the flame front as low-density burned gas compresses high-density unburned gas. Compression of the unburned gas increases its temperature and the fuel-oxidizer mixture detonates when a shock passes through the flame front through an energy-focusing mechanism. This detonation mechanism is consistent with previous work examining DDT in stoichiometric ethylene-oxygen mixtures \cite{GoodwinCNF2016,goodwin2017shock}.

Additional simulations were performed to investigate the sensitivity of the solution to initial conditions. Modifying the domain size and mesh resolution resulted in small changes to the time and location of initial detonation, but ignition in the boundary layer, growth of a Rayleigh-Taylor instability at the flame front, and detonation through an energy-focusing mechanism were observed in all cases. Future work will catalog the effect of inflow Mach number and fuel-oxidizer mixture composition on time to ignition, flow structure, suppression of fluid instabilities, and transition to detonation.

\section*{Acknowledgments}
The authors gratefully acknowledge the support of the Base Program at the Naval Research Laboratory and the Glenn L. Martin Institute Chaired Professorship at the A. James Clark School of Engineering, University of Maryland. The authors thank Professor Ryan Houim for the use of FAST (Flame Acceleration Simulation Tool). The computations were performed on a Department of Defense High Performance Computing Center system at the Air Force Research Laboratory.

\bibliographystyle{elsarticle-num-CNF}
\singlespacing
\bibliography{PROCI_2018_arXiv}


%
%
%
%
%
%
%
%
%
%
%
%
%

\end{document}